\documentclass[acmtog,screen, authorversion]{acmart}

\usepackage{booktabs} 
\usepackage{wrapfig}   
\usepackage{lipsum} 
\usepackage{soul}
\usepackage{bm}
\usepackage{xspace}
\usepackage[many]{tcolorbox}
\usepackage[normalem]{ulem}

\usepackage[noabbrev,capitalise,nameinlink]{cleveref}

\creflabelformat{equation}{#2\textup{#1}#3}%

\definecolor{orange}{rgb}{1, 0.5, 0.}
\definecolor{green}{rgb}{0, 0.6, 0}
\definecolor{purpleD}{rgb}{.8941, .8, .9043}
\definecolor{greenD}{rgb}{.7145, .9249, .7999}
\newcommand{\unsure}[1]{\textcolor{orange}{#1}}

\newcommand{\ichao}[1]{\textcolor{blue}{{#1}}}
\newcommand{\yuki}[1]{\textcolor{green}{{yuki: #1}}}
\newcommand{\ann}[1]{\textcolor{orange}{{#1}}}

\newcommand{\methodName}{GarmentImage\xspace}

\newcommand{\nonBoundary}{NON-BOUNDARY\xspace}
\newcommand{\nonStitch}{NON-STITCH\xspace}
\newcommand{\frontToBack}{FRONT-TO-BACK\xspace}
\newcommand{\sideBySide}{SIDE-BY-SIDE\xspace}

\newcommand{\etal}{{\it{et~al.}}}

\newcommand{\eg}{e.g.,}
\DeclareMathOperator*{\argmin}{arg\,min}

\tcbset{top=0mm,bottom=0mm,left=0mm,right=0mm,fonttitle=\bfseries\scriptsize\color{gray},colbacktitle=white,enhanced,attach boxed title to top center={yshift=-1mm},boxed title style={top=0mm,bottom=0mm,left=0mm,right=0mm}}
\acmJournal{TOG}

\copyrightyear{2025}
\acmYear{2025}
\setcopyright{acmlicensed}\acmConference[SIGGRAPH Conference Papers
'25]{Special Interest Group on Computer Graphics and Interactive Techniques
Conference Conference Papers }{August 10--14, 2025}{Vancouver, BC, Canada}
\acmBooktitle{Special Interest Group on Computer Graphics and Interactive
Techniques Conference Conference Papers (SIGGRAPH Conference Papers '25),
August 10--14, 2025, Vancouver, BC, Canada}
\acmDOI{10.1145/3721238.3730632}
\acmISBN{979-8-4007-1540-2/2025/08}

\begin{document}
\title{\methodName: Raster Encoding of Garment Sewing Patterns with Diverse Topologies}

\author{Yuki Tatsukawa}
\orcid{0009-0003-5128-8032}
\affiliation{%
 \institution{The University of Tokyo}
 \country{Japan}
}
\email{yukitatsu0817@gmail.com}

\author{Anran Qi}
\orcid{0000-0001-7532-3451}
\affiliation{%
 \institution{The University of Tokyo}
  \country{Japan}
}
\affiliation{%
 \institution{Centre Inria d'Université Côte d'Azur}
  \country{France}
}
\email{anran.qi@inria.fr}

\author{I-Chao Shen}
\orcid{0000-0003-4201-3793}
\affiliation{%
 \institution{The University of Tokyo}
 \country{Japan}
}
\email{ichaoshen@g.ecc.u-tokyo.ac.jp}

\author{Takeo Igarashi}
\orcid{0000-0002-5495-6441}
\affiliation{%
 \institution{The University of Tokyo}
 \country{Japan}
}
\email{takeo@acm.org}
\renewcommand{\shortauthors}{Yuki Tatsukawa, Anran Qi, I-Chao Shen, and Takeo Igarashi}

\begin{abstract}
Garment sewing patterns are the design language behind clothing, yet their current vector-based digital representations weren’t built with machine learning in mind. Vector-based representation encodes a sewing pattern as a discrete set of panels, each defined as a sequence of lines and curves, stitching information between panels and the placement of each panel around a body. 
However, this representation causes two major challenges for neural networks: discontinuity in latent space between patterns with different topologies and limited generalization to garments with unseen topologies in the training data. 
In this work, we introduce \methodName, a unified raster-based sewing pattern representation. 
\methodName encodes a garment sewing pattern’s geometry, topology and placement into multi-channel regular grids.
Machine learning models trained on \methodName achieve seamless transitions between patterns with different topologies and show better generalization capabilities compared to models trained on vector-based representation.
We demonstrate the effectiveness of \methodName across three applications: pattern exploration in latent space, text-based pattern editing, and image-to-pattern prediction.
The results show that \methodName achieves superior performance on these applications using only simple convolutional networks.
\end{abstract}

\begin{teaserfigure}
  \centering
  \includegraphics[width=\textwidth]{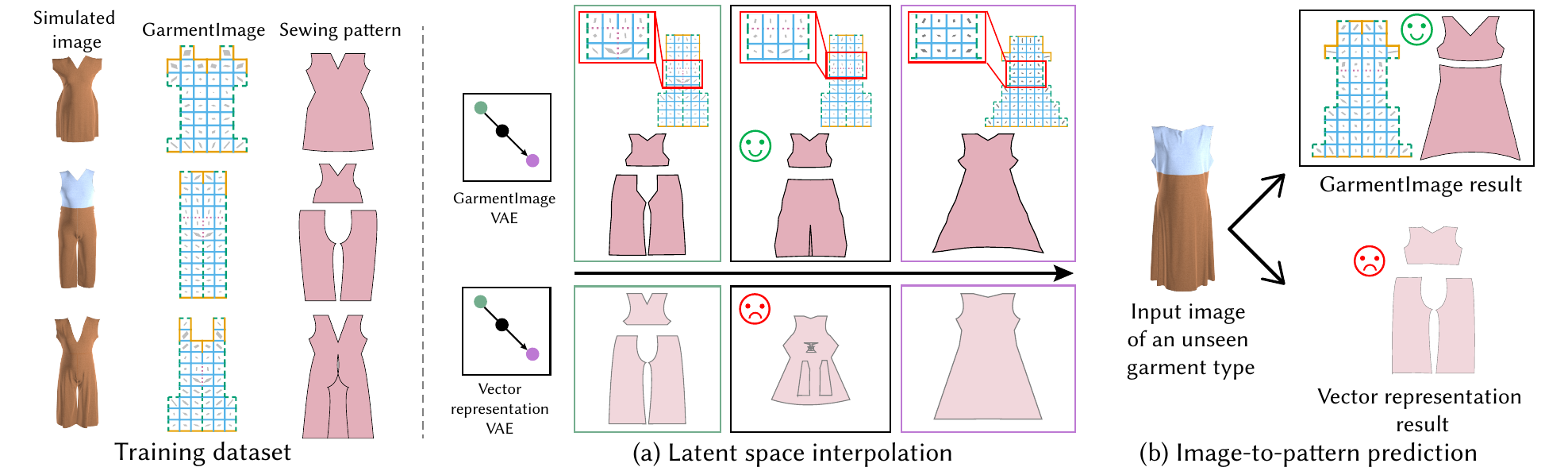}
  \caption{
  \methodName encodes a garment sewing pattern's geometry, topology and placement as raster data. This leads to a more continuous latent space and improved generalizability to unseen topologies compared to vector-based sewing pattern representation.
  (a) Interpolation between the two patterns with different topologies (green and purple) in the latent space of the GarmentImage-trained VAE yields a continuous transition and seamless panel merging (top), whereas the vector-based representation-trained VAE generates an invalid pattern (bottom).
  (b) When given an image of an unseen garment type (\textit{top + skirt}), the GarmentImage-trained model successfully predicts the new pattern (top), whereas the vector-based model defaults to a known pattern (\textit{top + pants}) present in the training data (bottom).
  }
  \label{fig:teaser}
\end{teaserfigure}

\ccsdesc[500]{Computing methodologies~Shape modeling}

\keywords{Garment pattern modeling, Garment pattern design}

\maketitle

\section{Introduction}
\label{sec:intro}

A sewing pattern is a pre-designed template to cut fabric pieces for creating a garment or textile item \cite{aldrich2015metric, armstrong2014patternmaking}. 
It typically consists of a collection of 2D fabric pieces, or \textit{panels}, with annotations specifying connectivity among borders and placements around a body. 
In the past, a sewing pattern was drawn directly onto fabric or paper using rulers and measuring tapes.
Later, in Computer-Aided Design systems, it is presented as a discrete collection of 2D shapes composed of lines and curves. 
In recent years, many learning-based methods have utilized this vector-based representation for various tasks, such as inferring garment sewing patterns from point cloud~\cite{Korosteleva2022}, photograph~\cite{liu2023towards}, and text~\cite{he2024dresscode}.
These advancements significantly simplify the process of designing garment sewing patterns.

Despite the widespread use of vector-based representation, we observe two challenges when using it in learning-based methods. \textbf{1. Discontinuity in latent space:} 
In vector-based representation, two patterns with similar overall 2D shapes but different topologies often have different numbers of panels and panel topologies. 
This causes the latent space created by encoding the vector-based representation to show significant discontinuities between patterns with different topologies (\autoref{fig:teaser}(a)).
\textbf{2. Limited generalization to garments with topologies unseen in the training data:} The above challenge also extends to pattern prediction from a given input, such as image or 3D model of a draped garment.
Visually similar inputs can lead to significantly different vector-based pattern representations.
This increases the Lipschitz bound of machine learning models trained to predict patterns from inputs, reducing their ability to generalize to patterns with unseen topologies (\autoref{fig:teaser}(b)). 
Additionally, handling variable numbers of panels across multiple garment types requires machine learning models to implicitly map input patterns to a discrete set of panels from a predefined panel pool.
This discrete set selection becomes particularly challenging for garments with unseen topologies, further limiting the generalization ability of a trained model.

 To address these challenges, we present \methodName, a novel raster representation of sewing patterns.
 \methodName integrates the discrete collection of 2D panels, the connectivity among panels, and the placement of panels into multi-channel 2D grids.
 Each grid cell contains an \textit{inside/outside} flag indicating occupancy, four edges with associated \textit{edge types} embedding stitching information,  and a local \textit{deformation matrix} capturing the panel geometry.   The placement of a panel around a body is implicitly represented by the location of its associated cells on the 2D grid.

\methodName  brings several benefits. 
First, the resulting representation is a 2D matrix of numerical values, similar to a raster image.
This enables  the use of established raster image generative modeling techniques to model, edit, and optimize sewing patterns. 
We employ simple convolutional neural networks to encode and decode the \methodName, demonstrating that even a straightforward network structure can effectively handle the complex pattern representation and its manipulation.
Second, changes in pattern topology are embedded within the grid structure in \methodName rather than as discrete panel selection in vector-based representation.
This leads to a continuous transition in the latent space between
patterns with different topologies, enabling both continuous topology interpolation and prompt-driven pattern optimization directly in the latent space (\autoref{fig:teaser}(a)).
Third, when inferring the pattern from an input, such as image, we can directly predict the \methodName as a whole and then procedurally reconstruct the discrete panel sets, effectively addressing the challenge of discrete panel selection.
Additionally, garments with similar visual looks have similar \methodName representations regardless of topology.
These properties improve the generalization ability of machine learning models trained on GarmentImage, leading to better pattern prediction performance on garments with unseen topologies (\autoref{fig:teaser}(b)).

To summarize, the contribution of this work is

\begin{itemize}
\item we present \methodName, a raster-based sewing pattern representation that integrates geometry, topology and placement information into multi-channel regular grids.
\item we showcase its advantages on interpolation and generalization ability over three applications: (1) VAE latent space exploration, (2) text-based pattern editing, and (3) image-to-pattern prediction.
\end{itemize}

\section{Related Work}
\label{sec:related}

\subsection{Garment Pattern Representation}
Garments in our daily lives are composed of various fabric pieces sewn together.
To translate this into a digital format, sewing patterns are often represented as a collection of polygons called panels, with stitching details defining how these panels are assembled.
The panel is often defined as a closed loop of parametric curves (e.g., B-splines or Bezier curves), defined by vertices and control points. 

In computer graphics, the processing of garment panels varies based on the nature of the task.
For tasks involving 3D surface data, such as garment simulation \cite{narain2012adaptive}, sewing pattern grading \cite{brouet2012design, wang2018rule} and sewing pattern adjustment from 3D shape editing \cite{bartle2016physics, qi2024perfecttailor} or sketch \cite{li2018foldsketch}, panels are often processed as a 2D mesh, which provides computational simplicity and ease of manipulation. 
For example, to simulate the details of the cloth, such as wrinkles, Narain et al. \shortcite{narain2012adaptive} presented each panel as a triangulated 2D mesh, and dynamically refined and coarsened triangle meshes to efficiently model the details.
On the other hand, for tasks that focus on the panel's shape only, e.g., garment pattern inferences from image \cite{su2020mulaycap, jeong2015garment, yang2018physics, liu2023towards, chen2024panelformer}, sketch \cite{Wang2018}, 3D point cloud \cite{Korosteleva2022}, a parametrized representation is usually adopted.
For example, to predict the sewing pattern of a given image, Yang et al. \shortcite{yang2018physics} parametrized the sewing pattern on several parameters that define the pattern size, then estimated those parameters with iterative optimization \cite{kennedy1995particle}.
Recently, GarmentCode~\cite{Korosteleva2023} and GarmentCodeData\cite{korosteleva2024garmentcodedata} presented a procedural way to generate the garment patterns at scale parametrized on body parameters, in which each garment panel is still represented by parametric curves. Design2GarmentCode \cite{zhou2024design2garmentcode} used fine-tuned Large Multimodal Models to directly generate GarmentCode~\cite{Korosteleva2023} programs from multi-modal design input.

Different from the aforementioned works, in this paper, we propose to model a garment pattern as raster data (2D grids), a bitmap representation composed of multiple concepts, including layers, inside/outside flags, edge types, and local deformation matrices \cite{Igarashi2005,Sorkine2007}.
Our work is inspired by methods in computer graphics that map 2D and 3D surfaces to a grid space.
Geometry Images represented a 3D surface into a 2D image \cite{Gu2002}, and it has been used to learn 3D surface models by neural networks \cite{Groueix2018, Sinha2016}.
Similarly, Polycube mapped a 3D surface to a set of 2D grid panels \cite{Tarini2004} and Polysquare mapped 2D shape onto a 2D grid \cite{xiao2018computing, liu2017distributed}. 
Due to diverse panel shapes and the equal seam length constraint, we use a simple strategy to robustly project panel shapes onto 2D grids. We also draw inspiration from Shen et al. \shortcite{shen2020gan}, which modeled the 3D garment as an image representation within the UV space of the human body. However, their work does not explore pattern representation. 


\subsection{Garment Pattern Prediction from an Input}

\noindent\textbf{Pattern prediction from 3D input.}
Given a 3D garment mesh as input, the shape is first segmented into patches guided by user sketches \cite{wang2005design}, predefined seams \cite{bang2021estimating}, or woven fabric properties \cite{Pietroni2022}. These patches are then parameterized into 2D using surface flattening techniques \cite{Sorkine2007, wang2002surface}. Recently, researchers have explored data-driven approaches for extracting sewing patterns from 3D data \cite{goto2021data, Korosteleva2022}.
Goto et al. \shortcite{goto2021data} computed the pattern classification on 3D input from multiple-image segmentation learned by an image translation network \cite{ronneberger2015u}.
NeuralTailor \cite{Korosteleva2022} represented a model that converts a 3D point cloud to a garment pattern, learning from a pattern data set \cite{Korosteleva2021}, where the model first generated a sequence of panels choosing from a set of predefined panel categories and then inferred their shape.

\noindent\textbf{Pattern prediction from 2D image.}
Given an image as input, a common approach is to match the garment with a parametric sewing pattern, then optimize the pattern's parameters for garment reconstruction from images \cite{su2020mulaycap, jeong2015garment, yang2018physics}. 
Wang et al. \shortcite{Wang2018} proposed to learn a shared latent shape space between  2D sketches, garment and body shape parameters, and draped garment shapes by training multiple encoder-decoder networks for each type of garment, enabling fast pattern inference from sketch. 
Recently, Liu et al. \shortcite{liu2023towards} created a comprehensive dataset with various human poses, body shapes, and sewing patterns, and introduce a two-level Transformer decoder to recover garment panels from learned panel queries defined for all panel types. 

\noindent\textbf{Pattern generation from text.} 
Recently, DressCode~\cite{he2024dresscode} proposed a method to generate sewing patterns from a text by first quantizing the pattern to a sequence of tokens and using a GPT-based architecture to generate the tokens autoregressively. 

These studies represent patterns as discrete panels, either using a set of parameterized Bézier curves or a sequence of tokens.
Thus, these approaches encounter the two challenges (discontinuity and less generalizability) outlined in \autoref{sec:intro}. 
In contrast, \methodName~represents a pattern as integrated raster data,  thus avoiding these challenges. We demonstrate its advantages on tasks such as pattern exploration in latent
space, text-based pattern editing and image-to-pattern prediction in~\autoref{sec:experiments}.

\section{\methodName}

\methodName representation encodes a sewing pattern---including a discrete collection of 2D panels, stitching information, and the placement of each piece on the body—as raster data (\autoref{sec:rep}). It serves as an intermediate data structure connecting vector-based pattern representation and machine learning models. Given a vector-format garment pattern, the encoding process (\autoref{sec:encoding}) transforms it into a GarmentImage suitable for input to learning-based methods. Conversely, once a GarmentImage is generated by a model, the decoding process (\autoref{sec:decoding}) reconstructs it into a vector-based pattern compatible with existing fashion pipelines, such as simulation. In what follows, we refer to a vector-based sewing pattern simply as a \textit{sewing pattern} or \textit{pattern}, and use the term \textit{GarmentImage} to denote our proposed sewing pattern representation.

\subsection{Representation}
\label{sec:rep}
\begin{wrapfigure}{r}{0.32\textwidth}
    \centering
    \includegraphics[width=0.31\textwidth]{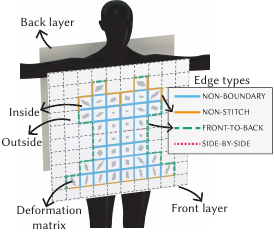}
    \caption{Representation overview.}
   \label{fig:representation_overview}
\end{wrapfigure}

\autoref{fig:representation_overview} provides an overview of the four core concepts in the GarmentImage representation. We describe each concept in detail below and explain how their corresponding values are computed in next section (\autoref{sec:encoding}).

\subsubsection{Layer}
\setlength{\columnsep}{1.3pt}
\setlength{\intextsep}{0.13pt}
GarmentImage represents a sewing pattern using two distinct layers: a front layer and a back layer. Panels positioned in front of the body are embedded in the front layer, while those behind the body are embedded in the back layer (front and back grid sandwich a T-posed body). Each layer is represented as a 2D array of grid cells.  Each grid cell is positioned in a specific location around the 3D human body. Adjacent grid cells are mapped to adjacent regions on the human body surface. Additionally, each grid cell contains an inside/outside flag, four edge types, and a deformation matrix, detailed below.

\subsubsection{Inside/Outside Flag}
By analogy to parts of the human body covered by a garment panel, we use the inside/outside flag of a cell to indicate whether the cell on the layer is covered by the garment panel or not.

\subsubsection{Edge Type}
Each grid cell has four edges, and each edge has a type that stores the boundary and stitching information. As illustrated in~\autoref{fig:representation_overview}, we define four edge types:
\begin{itemize}
\item \textbf{\nonBoundary}: The edge is not on a boundary or stitch. Completely inside a panel or outside a panel.
\item \textbf{\nonStitch}: The edge is on the boundary of a garment without stitching. This can appear inside a panel with holes.
\item \textbf{\frontToBack}: The edge is on the boundary of a panel and stitched to an edge at the same location on the opposite layer (front-to-back or back-to-front). 
\item \textbf{\sideBySide}: The edge is on the boundary of a panel and stitched to an edge of an adjacent cell on the same layer (front-to-front, back-to-back).  
\end{itemize}

\subsubsection{Deformation Matrix}

While each grid cell is associated with a local region on a garment panel, the shape of the grid cell (square) is different from the shape of the local region inside the panel (arbitrary quadrilateral).
The deformation matrix represents the mapping from the square to the quadrilateral.
More specifically, a deformation matrix \( F \in \mathbb{R}^{2 \times 4} \) consists of four columns, each of which represents a \textit{deformation vector} \( f \in \mathbb{R}^{2 \times 1} \) of an edge. It is defined as $\bar{v}_j-\bar{v}_i$, where $\bar{v}_i$ is the start of the edge and $\bar{v}_j$ is the end of the edge in the quadrilateral (edges are oriented).

\begin{figure*}[t]
  \centering
  \includegraphics[width=\textwidth]{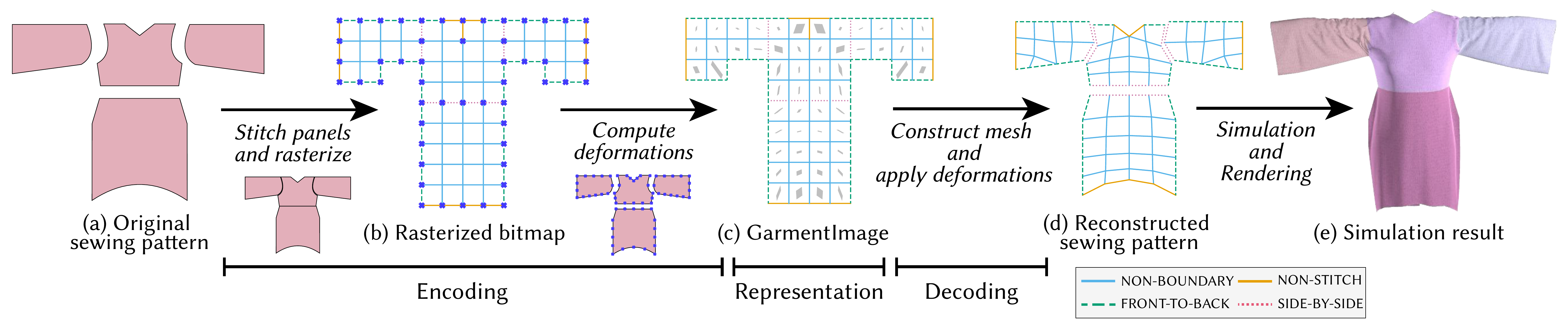}
  \caption{
  \textbf{\methodName encoding and decoding process.}
  A \methodName~is automatically encoded from a sewing pattern in vector format and can be decoded back to the vector format. Given a sewing pattern in vector format (a), we stitch the panels and fill the gaps between them before rasterizing it into a bitmap (b).  During this process, we establish correspondences between the original panel curves and the vertices on the bitmap grid. This allows us to assign edge types to each grid edge and compute deformation matrices that align the grid cells with the original panels. The resulting GarmentImage representation for each cell (c) contains inside/outside flags, edge types, and a deformation matrix, visualized as a parallelogram in the cell. For the decoding process, starting from the GarmentImage, we reconstruct the sewing pattern in vector format (d), which can then be used for applications such as simulation (e). 
 } 
  \label{fig:garmentImageConversion}
\end{figure*}

\subsection{Encoding}
\label{sec:encoding}
Equipped with the definition of GarmentImage representation described in \autoref{sec:rep}, the encoding process aims to convert the vector representation of a garment pattern into it.

Given a garment pattern, we first classify the panels into front panels and back panels using the panels' placement information.
If a panel covers both the front and back sides of the body, we need to split it into the front part and back part, which are then stitched together along the panel boundary. We then deform the panels so that panel seams are aligned with the corresponding seams to be stitched together (the intermediate layout between \autoref{fig:garmentImageConversion}(a) to (b)).
 Then, we embed the deformed panels into a grid by discretizing vertex coordinates.
 Each grid cell is classified as inside if its center lies within one panel, or outside if it lies outside all panels.
 As we have aligned seam edges between stitched panels before, this discretization can guarantee that (1) panels stitched together is adjacent in the grid space, (2) there is no gap or overlap between panels on the grid, and we can handle stitches between seams with different lengths (see the waistbands in \autoref{fig:complicated_garment}(b)).
 
 We then rasterize the front and back grids. After rasterization, GarmentImage can be intuitively considered as a front grid positioned in front of the human body and a back grid positioned behind it. The edge type is automatically determined by the relation between the adjacent cells (\autoref{fig:garmentImageConversion}(b) to (c)): the corresponding FRONT-TO-BACK edges are located in the same position at the front and back grids. A SIDE-BY-SIDE edge is an edge adjacent to another from a neighboring cell on the same grid, where the two cells belong to two switched panels.

The final step is to compute the deformation matrix for each grid cell (\autoref{fig:garmentImageConversion}(b) to (c)). We construct a quad mesh $M=(V, E)$ by connecting grid cells associated with a panel, where $V$ is mesh vertices and $E$ is the mesh edges. 
We then deform the quad mesh $M$ to $\bar{M}=(\bar{V}, \bar{E})$ so that its boundary matches the boundary of the original panel while minimizing the edge deformation. 
We formulate this as a least squares problem with a linear constraint \autoref{eq:encoding}: 
\begin{align}
     \argmin_{\bm{\bar{v}}} \{ {\sum_{(i,j)\in E}\!\bigl(\bigl(\bar{v}_j - \bar{v}_i\bigr) \;-\; \bigl(v_j - v_i\bigr)\bigr)^{2}} \}
    \quad \text{s.t.}
    \quad & C\,\bar{v} \;=\; c
    \label{eq:encoding}
\end{align}
where  $v_i$ and $\bar{v}_i$ are the positions of the $i$-th vertex of $M$ before and after deformation, $(i, j) \in E$ indicates  a directed edge in $M$ from vertex  $v_i$ to vertex $v_j$. The matrix $C$ encodes the linear constraint that enforces the boundary vertices in $\bar{M}$ to match their target positions on the panel boundary curve, given by $c$.
We solve \autoref{eq:encoding} using Lagrange multipliers \cite{golub2005cme}. The deformation vector is defined as $\bar{v}_j-\bar{v}_i$, and the deformation matrix for a cell is formed by stacking four such deformation vectors as columns.


\subsection{Decoding}
\label{sec:decoding}
Although deformation and discretization squash or stretch the original panel during encoding, the decoding process aims at reconstructing the sewing pattern (panels, stitches, and placement of panels) from \methodName that might be generated by a machine learning model (\autoref{fig:garmentImageConversion}(c) to (d)).
This process is deterministic and fully automated. We compute it in three steps as follows.


\noindent\textbf{Getting panels by cell clustering.} First, different panels need to be separated from \methodName.
We group grid cells flagged as inside and enclosed by boundary edges (i.e., grid cell edges with edge type \nonStitch, \frontToBack, \sideBySide). Each resulting group of grid cells corresponds to a distinct garment panel.

\noindent\textbf{Panel shape recovery.} Next, for each identified panel, we aim to recover its shape. 
We construct a quad mesh $M=(V, E)$ from the group of grid cells as that of in encoding, and deform the quad mesh $M$ to $\hat{M} = (\hat{V}, \hat{E})$.
The deformation aims to minimize the difference between resulting edge $\hat{e}_{i,j} = \hat{v}_j-\hat{v}_i$, and embedded deformation vector $f_{i,j}$ while keeping the location of the resulting mesh on the grid. 
We formulate it as another least squares problem: 
\begin{align}
 \argmin_{\bm{\hat{v}}} \{ {\sum_{i,j \in E}((\hat{v}_j-\hat{v}_i) - f_{ij})^2} +  \sum_{v_i \in V}(\hat{v}_i - v_i )^2 \}
\end{align}
where $v_i$ is the vertex position in $M$, $(i, j) \in E$ indicates  a directed edge in $M$ from vertex  $v_i$ to vertex $v_j$ and $f_{ij}$ is its encoded deformation vector.

\noindent \textbf{Stitching and placement recovery.} With the recovered panel shape, we extract stitching information by transferring edge types from the 2D grids to their corresponding deformed edges on the panel curves. This defines how different panels are connected. Additionally, the location of the grid cells associated with a panel indicates the panel’s intended placement around the human body. This spatial information enables the reconstruction of a 3D garment mesh around a 3D human model for physical simulation, or the generation of 2D printable patterns that can be cut and stitched to produce a physical garment.

\subsection{Examples}
\begin{figure}[t]
  \includegraphics[width=\linewidth]{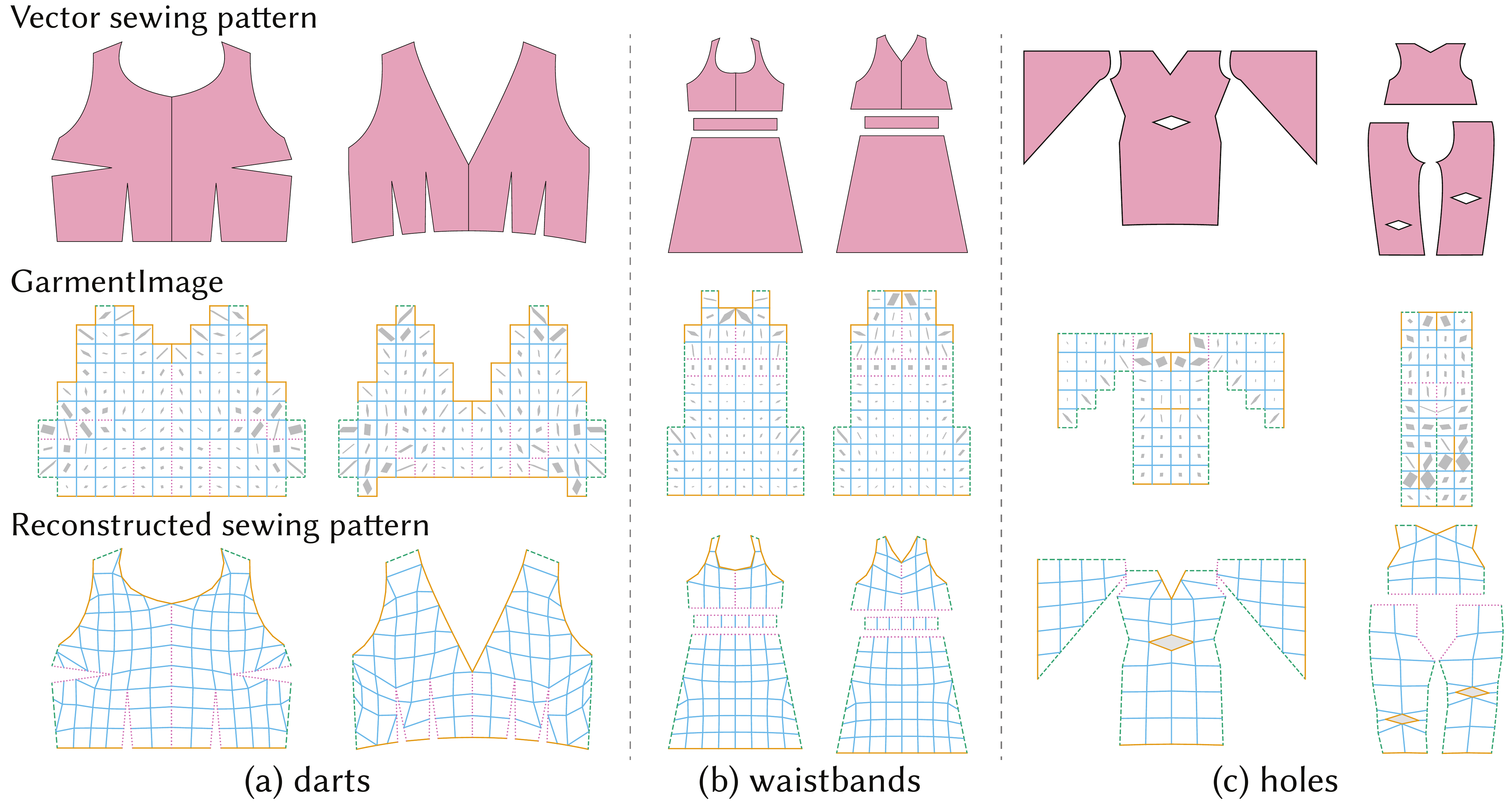}

  \caption{
  \textbf{\methodName examples.}
  \methodName can handle diverse garments and features such as darts (a), waistbands (b), and holes (c). 
}
  \label{fig:complicated_garment}
\end{figure}

In \autoref{fig:complicated_garment}, we show that our representation can represent a large variety of sewing patterns and design features such as waistbands and darts. 
Additionally, \methodName can naturally represent panels with holes, a capability that would require a dedicated template \cite{Korosteleva2021, Korosteleva2022} or command \cite{korosteleva2024garmentcodedata} in vector-based representation.

\section{Experiments}
\label{sec:experiments}
In this section, we introduce three downstream tasks to illustrate the advantages of \methodName.
Across all experiments, we use a $16 \times 16$ \methodName representation with $34$ channels per grid cell, resulting in a shape of $(34, 16, 16)$.
Of these $34$ channels, $17$ are allocated to the front layer and the remaining $17$ to the back layer. Each set of 17 channels includes one channel for the inside/outside flag, eight channels encoding the edge types for the bottom and left edges (as two one-hot vectors of length four), and eight channels representing the deformation matrix. To build our experimental datasets, we first sampled diverse sewing patterns by randomizing values of various parameters (\eg~sleeve length) defined in sewing pattern dataset  Korosteleva~\etal~\shortcite{Korosteleva2021}. Corresponding GarmentImages representations were then automatically generated from these patterns by the encoding process introducted in \autoref{sec:encoding}. In addition, we employed the simulator proposed in Korosteleva~\etal~\shortcite{korosteleva2024garmentcodedata} to collect simulated images from sewing patterns.

\subsection{VAE Latent Space Exploration}
\label{sec:vae}
We hypothesize that GarmentImage enhances the robustness of a trained VAE to new topologies, leading to a more continuous latent space than that of vector-based representation.
To validate this, we train a VAE using GarmentImage (dubbed GarmentImage VAE) and another using a vector-based pattern representation (vector-based VAE) on the same reconstruction task. We compare their performance based on latent space interpolation and extrapolation results.

Since \methodName~is a raster format, we can take advantage of popular neural network architectures such as convolutional neural networks (CNNs), which are well-suited for raster data.
We train a CNN-based VAE with $32$ dimension latent space using \methodName.
For the vector-based representation, we adopt the similar pattern format proposed in \cite{Korosteleva2021}. It consists of $4 \times n \times 16$ dimensions, where $4$ corresponds to (x, y, curvature x, curvature y), $n$ is the number of panel types, and $16$ is the maximum number of vertex in a panel.
To process this format, we employ a transformer-based model like Sewformer~\cite{liu2023towards}.
Specifically, we use a transformer-based VAE architecture inspired by Motionformer~\cite{petrovich2021action}, which injects two special tokens into the input sequence to predict the mean $\mu$ and standard deviation $\sigma$\ of the latent presentation. See~\autoref{fig:VectorRepresentationVAE} for more details.

We prepare three datasets. 
The first dataset (\autoref{fig:VAEInterpolation}(a,b)) includes three garment pattern types--\textit{dresses}, \textit{jumpsuits}, and \textit{top + pants}, and serves to assess whether the trained VAE can generate an entirely new garment pattern with new topology (\textit{two-panel dresses}) through latent-space exploration. The second dataset (\autoref{fig:VAEInterpolation}(c,d)) contains a wider variety of patterns, such as \textit{one-panel dress with sleeves} and \textit{top + skirt with sleeves}, and is used to evaluate the smoothness of the latent spaces when performing latent space interpolation across diverse garment pattern types.
The third dataset (\autoref{fig:VAEInterpolation}(e, f)) consists of \textit{one-panel shirts} and \textit{two-panel shirts with} and \textit{without darts}, and is used to test latent space interpolation of \methodName~in the presence of the complicated garment feature. We collected around 20,000 garment sewing patterns and GarmentImages for each garment type.

\noindent\textbf{Latent space interpolation.} 
As shown in \autoref{fig:VAEInterpolation}(a,c,e), the latent space of \methodName~VAE exhibits continuous interpolations between garments of different topologies.
For example, when interpolating between a \textit{top + pants} and a \textit{one-panel dress} pattern (\autoref{fig:VAEInterpolation}(a)), the topology of generated pattern transitions continuously in the \methodName~VAE.
In contrast, the vector-based VAE produces discrete jumps, often resulting in invalid garment patterns—an undesirable behavior for latent space interpolation. Additionally, our method can also support a continuous interpolation in the number and size of darts, as shown in \autoref{fig:VAEInterpolation}(e).

\noindent\textbf{Latent space extrapolation.} 
We assess the model’s ability to generate patterns with unseen topologies through the latent space extrapolation experiments. 
As shown in~\autoref{fig:VAEInterpolation}(b), when we aim to perform a topology edit -- applying the latent vector difference from a \textit{jumpsuit} to \textit{top + pants} patterns to a \textit{dress}, \methodName~VAE successfully splits the \textit{dress} into top and skirt panels, while the vector-based VAE fails to capture this transfer. In ~\autoref{fig:VAEInterpolation}(d), \methodName~VAE can even generate an unseen pattern (\textit{top + skirt}) purely by the topology edit in the latent space. In ~\autoref{fig:VAEInterpolation}(f),
GarmentImage VAE successfully transfers the adding darts edit. These findings demonstrate that \methodName~VAE's latent space remains continuous and robust to topological changes, even for garments not encountered during training.



\subsection{Text-based Pattern Editing}
\label{sec:prompt-to-pattern}
\begin{figure}[!t]
  \centering
  \includegraphics[width=\linewidth]{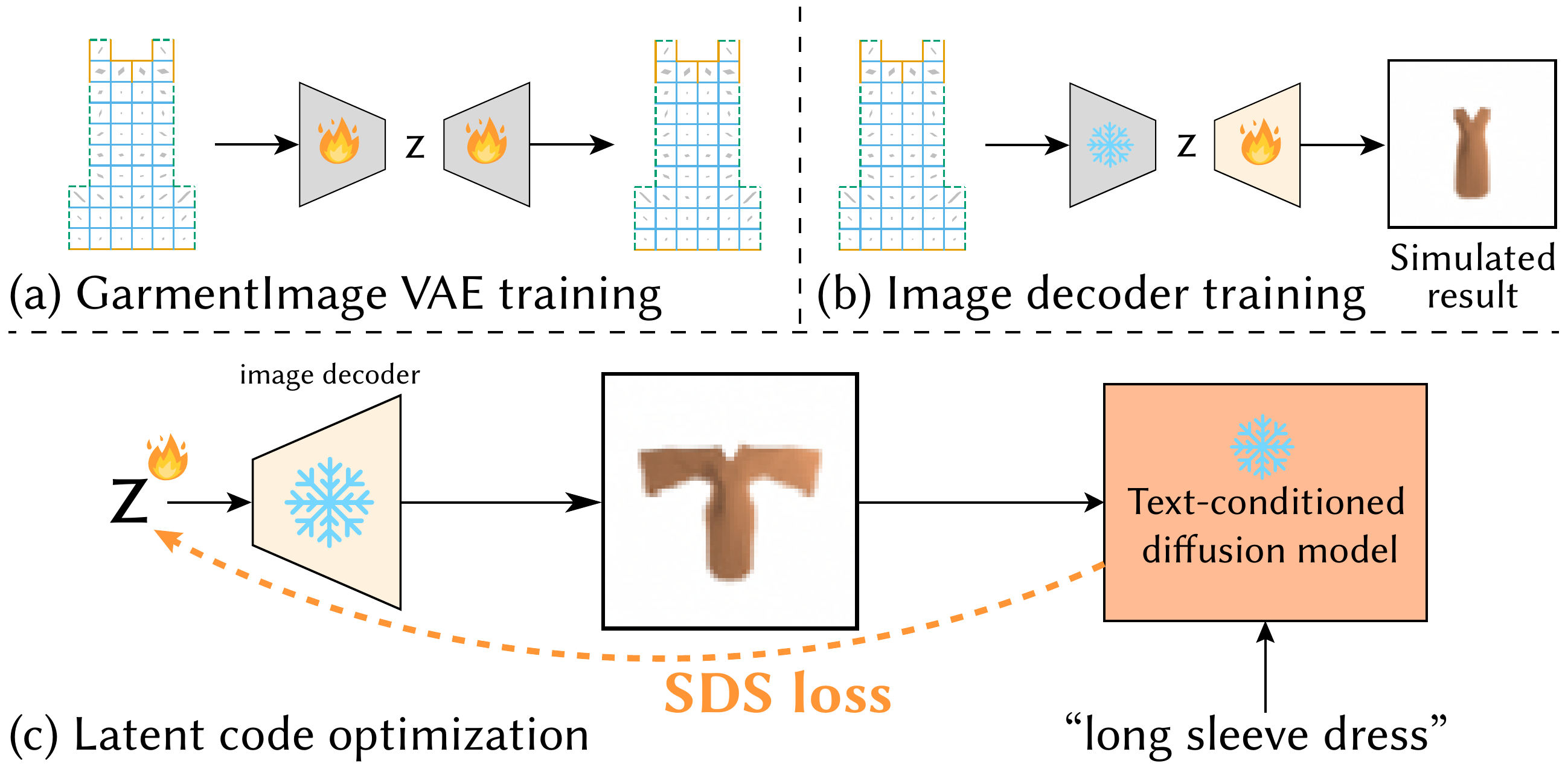}
  \caption{
  \textbf{Text-based pattern editing pipeline.} (a) We first train a \methodName VAE encoder and decoder on the \methodName reconstruction task.
  (b) Then we train an image decoder that predicts a simulated result in $64\times64$ given a latent code from  \methodName VAE.
  (c) We optimize for the best \methodName VAE latent code that minimizes the SDS loss ~\cite{pooledreamfusion} to conform to the input text prompt.
  }
  \label{fig:prompt-to-pattern-pipeline}
\end{figure}

\begin{figure}[ht]
  \centering
  \includegraphics[width=\linewidth]{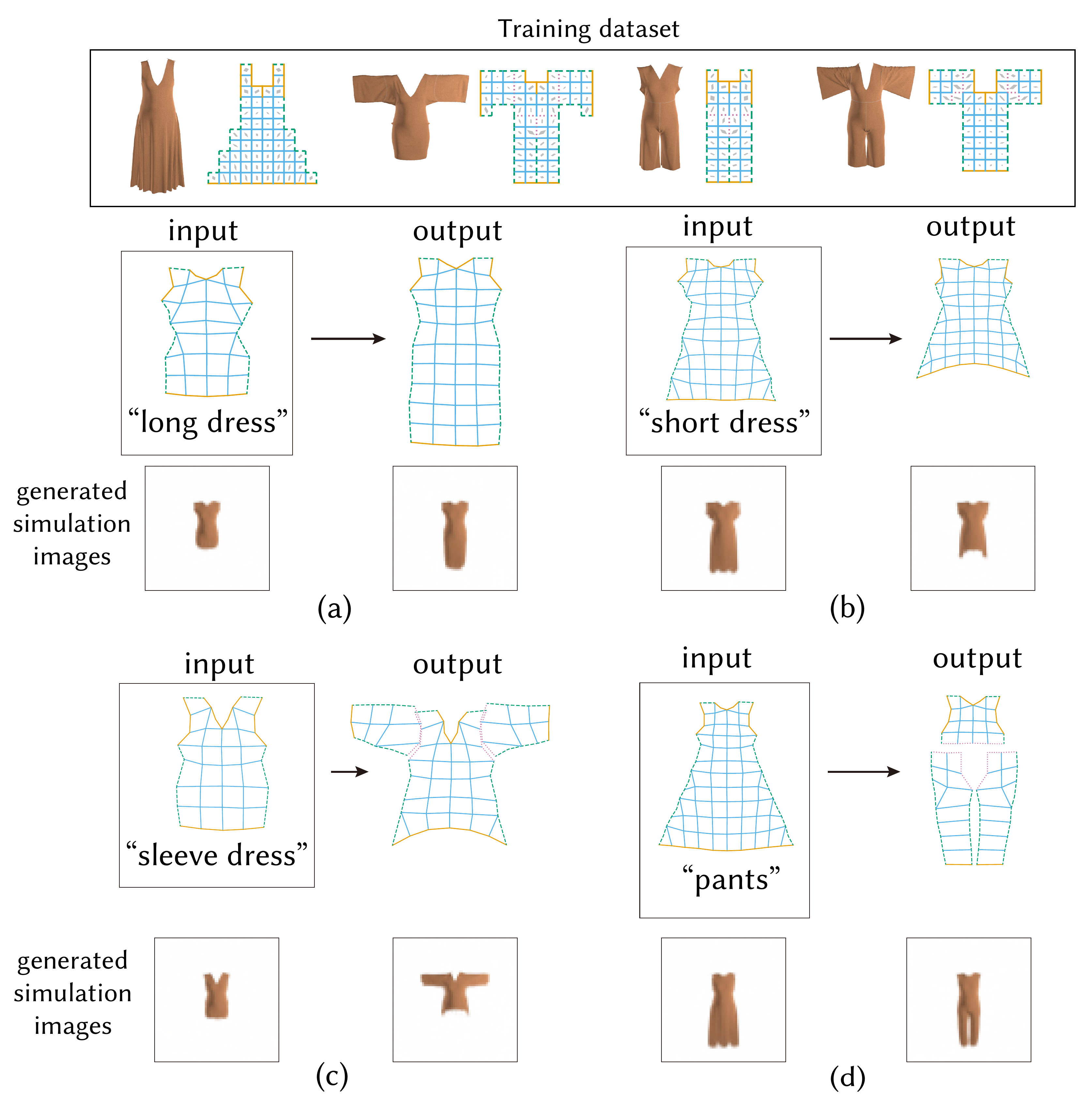}
  \caption{\textbf{Text-based pattern editing results.}
The input \methodName patterns adjust their geometry (a, b) and even topology (c, d) to match the input text prompts.
 }
  \label{fig:prompt-to-pattern-result}
\end{figure}

We demonstrate an optimization-based text-based pattern editing application, using the \methodName~VAE latent space described in \autoref{sec:vae}. In \autoref{fig:prompt-to-pattern-pipeline}, we optimize the latent code within this space to align with a given text prompt. We use Stable Diffusion\footnote{\url{https://huggingface.co/stable-diffusion-v1-5/stable-diffusion-v1-5}} \cite{rombach2022high} as the text-to-image generator. To bridge the domain gap with Stable Diffusion, we employ an image decoder that maps the VAE latent code to a simulated garment image. Finally, we utilize the SDS loss~\cite{pooledreamfusion} to minimize the distribution gap in the diffusion model's latent space between the generated simulation image and the target text prompt. 
We collected a dataset with paired \methodName and simulation result image for four garment types: \textit{one-panel sleeveless dresses}, \textit{one-panel dresses with sleeves}, \textit{tops + pants}, and \textit{tops with sleeves + pants}.
For simplicity, all simulated garments are rendered from the front view in brown, at a resolution of $64 \times 64$.
We train the image decoder on these paired samples while keeping the VAE encoder fixed. As shown in \autoref{fig:prompt-to-pattern-result}, the initial GarmentImages adapt their shapes and topologies to match the target prompts, thanks to the continuous latent space.
For instance, in \autoref{fig:prompt-to-pattern-result}(d), the \textit{one-panel dress} transforms into a \textit{top + pants} garment, while preserving the shape of the original top's silhouette, based on the text prompt "pants".

\subsection{Image-to-Pattern Prediction}
\label{sec:image-to-pattern}

In this experiment, we compare a model based on \methodName~with Sewformer~\cite{liu2023towards} and NeuralTailor \cite{Korosteleva2022} on image-to-pattern prediction task. Our goal is to demonstrate the generalization advantages of our representation against new garment topologies.

\subsubsection{Network Architecture}
\autoref{fig:unet_arch} illustrates our use of a simple U-Net model to sequentially predict a GarmentImage. First, the model predicts inside/outside flags from the input image. Next, it predicts edge types using both the input image and the previously predicted flags. Finally, the deformation matrix is predicted based on the input image, inside/outside flags, and edge types. We use L2 loss for the inside/outside flag and deformation matrix prediction, and cross-entropy loss for edge type prediction.


\subsubsection{Training and Evaluation}
\label{sec:training_and_result}

We evaluate the model's generalizability to unseen garment types in two experimental settings.

\noindent\textbf{Generalizability to an entirely new topology.} In this experiment, we evaluate whether the trained models can generalize to a garment topology that was never encountered during training (\autoref{fig:image-to-pattern} left).
To this end, we constructed a dataset D1 including four garment types: \textit{one-panel dresses}, \textit{one-panel jumpsuits}, \textit{top + pants}, and \textit{top + skirt}.
Each entry in D1 includes a sewing pattern, its \methodName~representation and simulated image.
The top panels are rendered in light blue, while the others are brown, clearly indicating the separation between the top and bottom panels.
We train the model using only \textit{dresses}, \textit{jumpsuits}, and \textit{top + pants} garments. 
We test the model on \textit{top + skirt} garments to assess how well the model generalizes to this entirely new topology.
We collected approximately $80,000$ garments for training. We compare our approach with Sewformer~\cite{liu2023towards}.
\autoref{fig:image-to-pattern} (left) shows that our model accurately predicts separate skirt panels for the garment's bottom, while Sewformer is limited to generating only pants panels or invalid ones.
We attribute Sewformer's limitations to its neural network architecture, designed to output vector-based sewing pattern representations. For each garment type, Sewformer implicitly performs panel selection from predefined panel pools, restricting its flexibility and making it difficult to generalize to entirely new garment topologies. In contrast, \methodName~encodes garment topology through a combination of layers, inside/outside flags and edge types, enabling it to handle new garment configurations without needing prior definitions.

\noindent\textbf{Generalizability to an unseen panel combination.}
In this experiment, we investigate whether the trained model can generalize to a garment panel combination not presented in the training data (\autoref{fig:image-to-pattern} right).
To this end, we constructed a dataset D2 with four garment types, \textit{one-panel dresses}, \textit{dresses with sleeves}, \textit{top + pants}, and \textit{top with sleeves + pants}.
To distinguish pants from skirts, we render the pants in light blue.
We train the model using only \textit{one-panel dresses}, \textit{dresses with sleeves}, and \textit{top + pants} garments. And evaluate on \textit{top with sleeves + pants} garments—an unseen combination of panels. Note that while each individual panel (top, sleeves, and pants) is present in the training data, their combination in this form is not. We compare our approach with Sewformer \cite{liu2023towards}.  Our training dataset contains approximately $80,000$ garments, while the test set includes around $6,500$ garments of seen types and $3,500$ of the unseen type. \autoref{fig:image-to-pattern} (right) shows that our model correctly predicts the new panel combination, while Sewformer fails to generate the sleeves and even valid patterns.

We also compute the intersection-over-union (IoU) between the predicted sewing patterns and ground truth patterns for each panel aligning their centroids for a fair comparison.  \autoref{tab:imageToPattern} presents the results alongside  NeuralTailor \cite{Korosteleva2022} and Sewformer. For NeuralTailor, we adapt the model to accept image inputs by replacing its point cloud encoder with a pre-trained ResNet-50 \cite{he2016deep}. This modified encoder extracts both per-pixel features and a global image representation, which are then fed into NeuralTailor’s original decoder to generate sewing patterns. The results indicate that our model demonstrates superior generalizability to the unseen panel combination compared to both NeuralTailor and Sewformer. 

\begin{table}[]
\centering

\caption{
We report the IoU values between the predicted sewing patterns and ground truth for both seen and unseen panel combinations on dataset D2. For seen panel combinations, our approach outperforms NeuralTailor \cite{Korosteleva2022} by $9.1\%$ and achieves performance comparable to Sewformer \cite{liu2023towards}.
Notably, for the unseen panel combination, our approach significantly surpasses both baseline methods.}

\begin{tabular}{lcc}
\toprule
                 & Seen combinations          & Unseen combination          \\ \midrule
NeuralTailor   & 0.8390                      & 0.3131         \\
Sewformer       & \textbf{0.9311}             & 0.4260          \\
Ours             & 0.9304                      & \textbf{0.8127} \\ \bottomrule
\end{tabular}
\label{tab:imageToPattern}
\end{table}

\section{Limitations and Future Work}
\label{sec:limitations}
While our method performs well in several applications, we acknowledge that there is significant room for improvement before it can be widely embraced by the fashion industry.

\textit{Non-smooth pattern boundaries.} A reconstructed sewing pattern from \methodName may appear distorted or less smooth compared to its original vector-based representation. 
This limitation can be mitigated by increasing the resolution of \methodName and incorporating post-processing to smooth the reconstructed patterns.

\textit{Non-uniqueness of the representation.} 
The same pattern can be encoded into different GarmentImages. For example, a panel may be represented as either a dense arrangement of many small cells or a sparse configuration of fewer large cells. To minimize inconsistencies and potential negative impacts on model performance, it is important to follow a consistent policy when converting patterns into GarmentImages for training.

\textit{Limitation of the automatic encoding.}
In our experiments, we constructed the GarmentImage datasets by automatically encoding vector-based patterns.
However, the current automatic encoding process is limited to simpler garment types, such as \textit{dresses} and \textit{jumpsuits}, and may produce invalid GarmentImages (see Figure 7(a)). The success rate for automatic encoding is $89.88\%$ for \textit{dresses} in the NeuralTailor dataset~\cite{Korosteleva2022}. We filtered out the unsuccessful examples from the training dataset using simple filtering rules, without requiring human intervention.
Handling more complex garments and addressing corner cases may require further processing or manual annotation.

\begin{figure}[!t]
  \centering
  \includegraphics[width=0.95\linewidth]{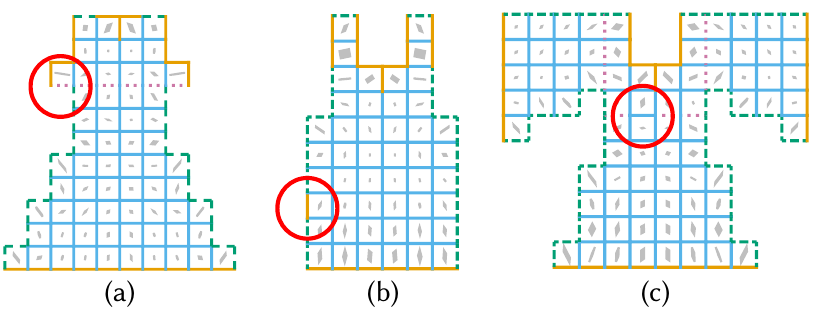}
  \caption{
  (a) \textbf{Invalid encoded \methodName.}
  Our automatic encoding process may generate invalid \methodName representations.
  In this case, the grid cell containing the \sideBySide edge should be marked as inside, but it is not.
  (b, c) \textbf{Invalid predicted \methodName.}
  Our proposed neural network may predict invalid \methodName representations.
  In (b), the \nonStitch edge should be predicted as \frontToBack, and in (c), the \nonBoundary edge should be predicted as \sideBySide.}
  \label{fig:invalidGarmentImage}
\end{figure}

\begin{figure*}[h]
  \centering
  \includegraphics[width=\linewidth]{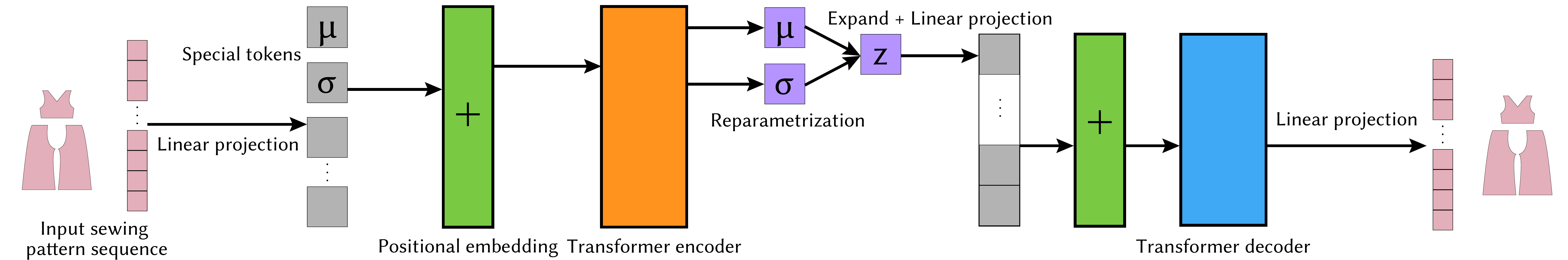}
  \caption{
  \textbf{Network architecture of vector-based VAE.}
  We employ a transformer-based architecture that takes a sewing pattern sequence and two special tokens ($\mu$ and $\sigma$ tokens) as input.
  Specifically, we give the input tokens into the transformer encoder to produce the mean$\mu$ and standard deviation $\sigma$ of the latent presentation.
  We then compute $z$ by reparametrization trick and pass it into the transformer decoder to reconstruct the input sewing pattern sequence.
  Note that this VAE only predicts the panels' shape and does not output stitch information.
  } 
  \label{fig:VectorRepresentationVAE}
\end{figure*}

\begin{figure*}[h]
  \centering
  \includegraphics[width=\linewidth]{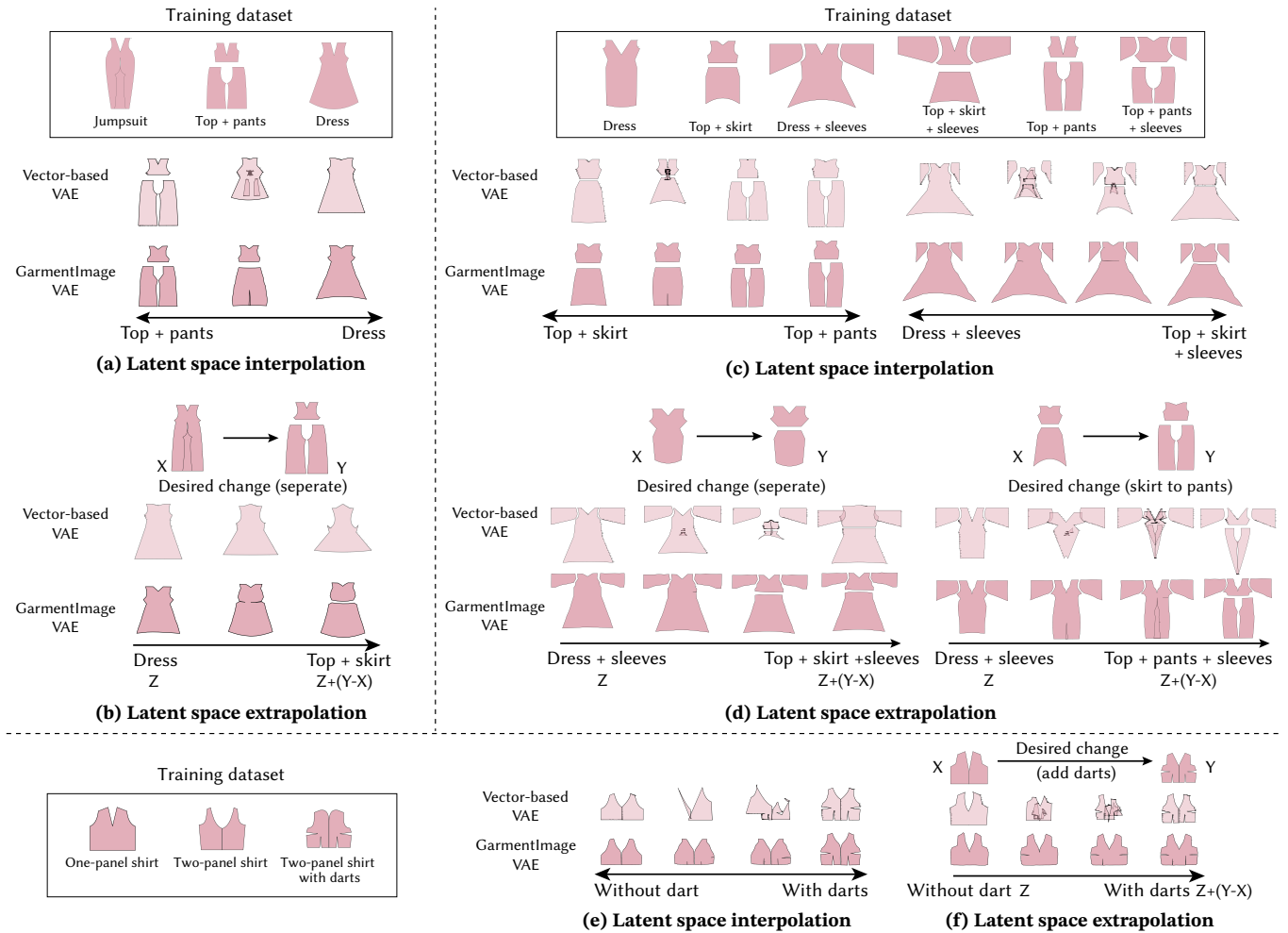}
  \caption{\textbf{Latent space experiment results.} The latent space of \methodName VAE showcases more continuous interpolations and extrapolations between different topologies than that of vector-based VAE. In (c, left), when interpolating a \textit{Top + skirt} into \textit{Top + pants} pattern, \methodName VAE continuously splits the skirt panel into two panels for pants, whereas the vector-based VAE exhibits discrete jumps. A similar behavior is observed in (a). In (e), our method achieves a continuous interpolation in the number and size of darts, while the vector-based VAE fails to generate valid dart designs. In (b), we demonstrate a topology edit by transferring the latent vector difference from a \textit{one-panel jumpsuit} to \textit{top + pants} pattern, and applying it to a \textit{one-panel dress}. \methodName~VAE successfully transfers the edit, splitting the \textit{one-panel dress} pattern into top and skirt panels. In contrast, the vector-based VAE fails to capture the intended transfer. Furthermore, in (d) \methodName~VAE can generate a previously unseen pattern—a \textit{two-panel dress}—purely by latent space extrapolation. In (f), \methodName~VAE successfully transfers the adding darts edit onto a \textit{one-panel shirt}. The vector-based VAE fails in both cases.}
  \label{fig:VAEInterpolation}
\end{figure*}

\begin{figure*}[ht]
  \centering
  \includegraphics[width=\textwidth]{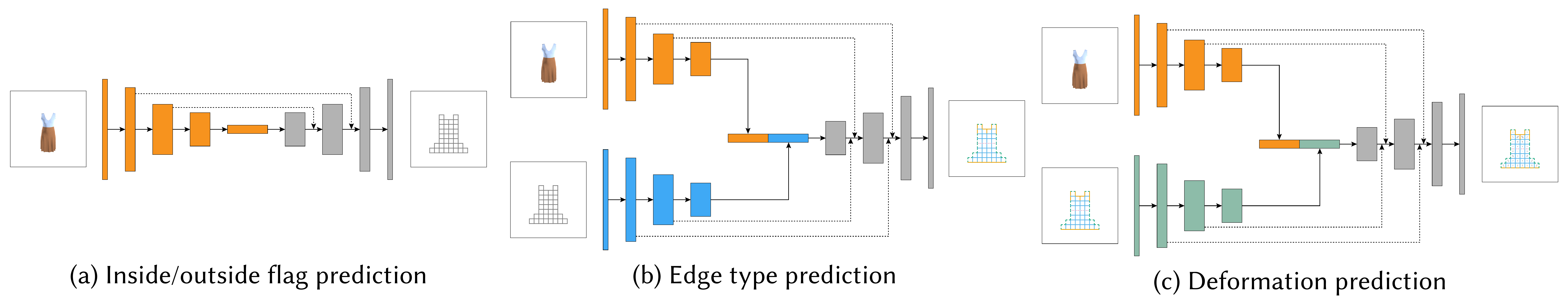}
  \caption{\textbf{Network architecture for Image-to-Pattern prediction (\autoref{sec:image-to-pattern}).}
  We employ three separate U-Net models in a step-by-step pipeline.
  Given an input image, (a) the first U-Net predicts the inside/outside flags.
  (b) Using the same input image and the predicted inside/outside flags, the second U-Net infers the edge types.
  Finally, (c) the third U-Net predicts deformation matrices from the input image and the previously generated inside/outside flags and edge types.
  Each model is trained independently using its ground truth intermediate representations. During inference, these models are applied sequentially.
 }
  \label{fig:unet_arch}
\end{figure*}

\begin{figure*}[ht]
  \centering
  \includegraphics[width=\textwidth]{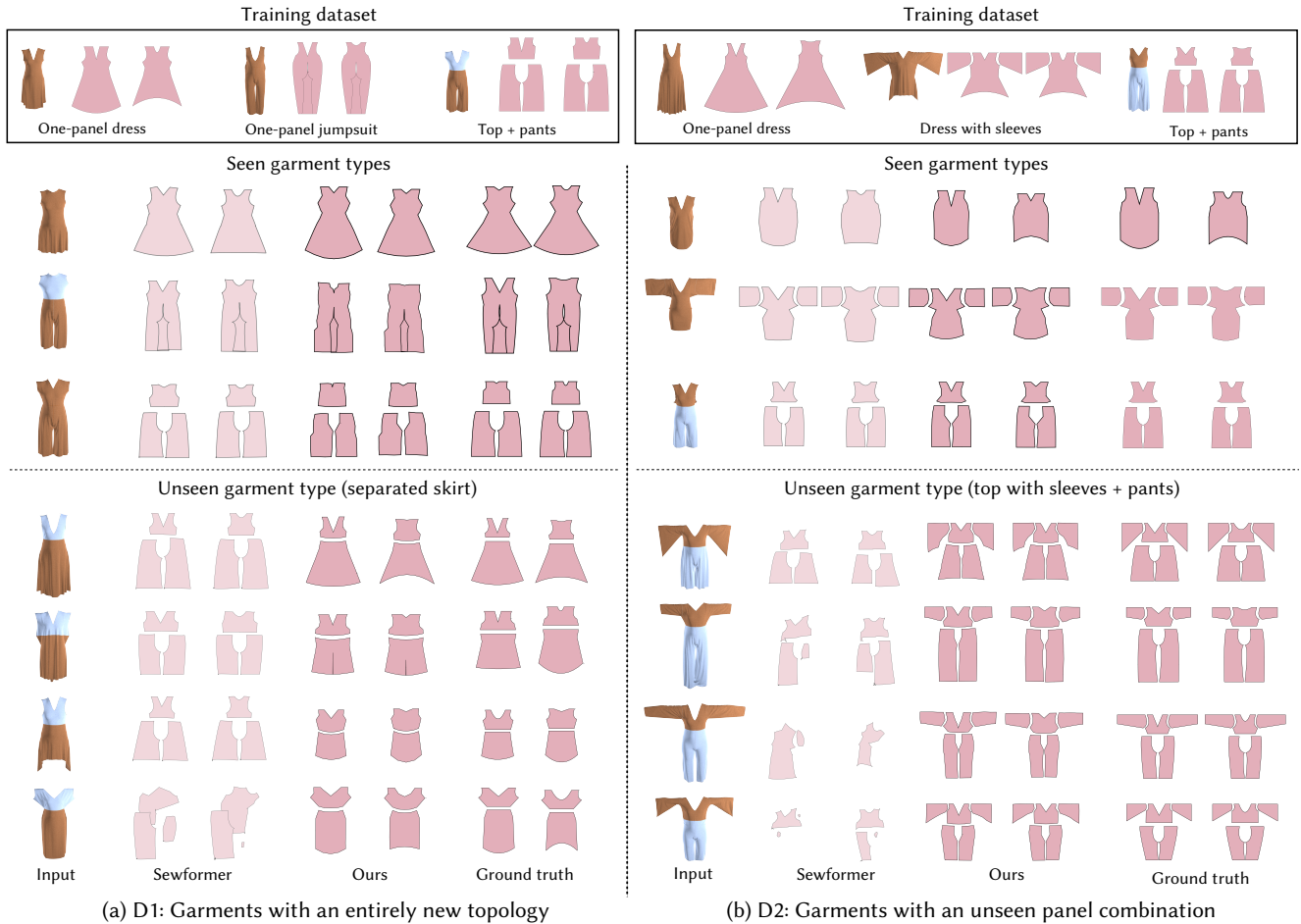}
  \caption{\textbf{Image-to-Pattern prediction results (\autoref{sec:image-to-pattern})}. We visualize the predicted patterns from an input image using our method and Sewformer \cite{liu2023towards}. While both methods can predict reasonable sewing patterns for garment types included in the training dataset (top), our method demonstrates superior generalizability for unseen garment types (bottom). For garments with an entirely new topology (a), our method predicts accurate patterns, whereas Sewformer usually defaults to patterns presented in the training data. For garments with unseen panel combination (b), Sewformer often produces invalid patterns, and it frequently omits sleeve panels, despite that the sleeve panel is clearly included in \textit{dress with sleeves} in the training dataset. In contrast, our method accurately predicts the correct garment patterns. }

  \label{fig:image-to-pattern}
\end{figure*}

\textit{Invalid predicted \methodName result.}
\methodName representations generated by neural networks might yield invalid garment patterns after decoding.
We occasionally encounter invalid outputs, often due to incorrect edge type predictions.
We automatically detect two issues: a \nonStitch edge between two \frontToBack edges (\autoref{fig:invalidGarmentImage}(b)), and a \nonBoundary edge between two \sideBySide edges (\autoref{fig:invalidGarmentImage}(c)). These errors are corrected by reassigning the appropriate edge types. However, fully resolving more complex inconsistencies may require human intervention.

\textit{Unrealistic garment dataset.}
Currently, the experiments in~\autoref{sec:prompt-to-pattern} and~\autoref{sec:image-to-pattern} utilize front-view renderings of simulated garments. While this setup effectively showcases the superior generalizability of our method, incorporating more realistic images will be essential for future work aimed at practical, real-world applications.


\textit{Extend \methodName to more diverse garment features.} In this work, we introduce the basic \methodName, which comprises two layers corresponding to the front and back sides of a garment. While layered garment features such as collars, cuffs, and pockets can be defined as separate panels on the new layer, tuck requires a distinct edge type labelled as \textbf{TUCK}.

\section{Conclusion}
In this work, we presented \methodName, a novel raster representation for diverse garment sewing patterns. 
\methodName encodes geometry, topology, and placement information into multi-channel grids, providing a unified alternative to traditional vector-based pattern representations. We demonstrated the advantages of \methodName using three applications: VAE latent space exploration, text-based pattern editing and image-to-pattern prediction.
Our results show that models trained with \methodName exhibit a more continuous latent space and improved generalization to unseen topologies, compared to vector-based pattern representations. Looking further, we hope that our work will inspire research exploring the use of the proposed representation in various garment design applications.



\begin{acks}
We thank Nicolas Rosset for the advice on VAE experiments and the anonymous reviewers for their valuable feedback.
This work was supported in part by the Japan Science and Technology Agency (JST) as part of Adopting Sustainable Partnerships for Innovative Research Ecosystem (ASPIRE), Grant Number JPMJAP2401, and JSPS Grant-in-Aid JP23K16921, Japan.
\end{acks}
%
\bibliographystyle{ACM-Reference-Format}
\bibliography{paper}

\clearpage

\end{document}